# LES of Turbulent Flow of The Non-Newtonian Fluid: The Turbulence Statistics

Mohamed Abdi [1], Meryem Ould-Rouiss [2] Abdelkader Noureddine [1]

[1] Laboratoire de Mécanique Appliquée, Université USTO Mohamed Boudiaf, BP 1505 Oran EL M'Naouar, Oran, Algeria

[2] Laboratoire de Modélisation et Simulation Multi Echelle, MSME, Université Paris-Est, UMR 8208 CNRS, 5 bd Descartes, 77454 Marne-la-Vallée, Paris, France

E-mail : abdi.mohamed1@live.fr

*Abstract*

*A large eddy simulation (LES) with an extended Smagorinsky model has been carried to investigate numerically the fully developed turbulent flow of a shear thinning fluid (n=0.75) in a stationary pipe at a simulation's Reynolds number equals to 4000 with a grid resolution of $65^3$ gridpoints in axial, radial and circumferential directions and a domain length of 20R. The present study set out to critically evaluate the influence of the Non-Newtonian rheological and hydrodynamic behaviour on the turbulence main features, as well as to ascertain the accuracy and reliability of the laboratory code predicted results. The turbulent flow statistics obtained compared reasonably well with the experimental data and DNS results. A reasonably good agreement has been obtained between the predicted results and available results of literature. The main findings suggest that the mean axial velocity fluctuations were generated in the wall vicinity and transferred to the radial and spanwise components, which resulted in to an enhancement in the kinetic energy of turbulent fluctuations along the radial direction. The energy transport from the axial fluctuations to the other fluctuations components between the flow layers was suppressed compared to that of the Newtonian fluid.*

## 1. Introduction

The Non-Newtonian fluids has received an increased attention across a number of disciplines both from academic and practical points of view over the recent decades, this kind of fluids is of great importance in mechanical and engineering fields and is encountered in a variety of engineering applications such as petroleum, food products, plastics and pharmaceuticals. The fully developed turbulent flow on the Non-Newtonian fluids in a straight smooth axially pipe has been received a lot of intention in the last decades by many researchers. Recently, there has been extensive research performed
in order to grasp and understand the rheology, hydrodynamic and thermally behaviour either experimentally [1-5] by or numerically by [6-16].

Rudman et al [9] employed a spectral element-Fourier method to report numerically Direct numerical simulation

(DNS) of a fully developed turbulent flows of power-law fluids (shear thinning) in a cylindrical stationary pipe, with domain length of 4-8πD for flow index flow of $n$=0.5, 0.69 and 0.75. At three simulation Reynolds number of 4000, 8000 and 12000 and over flow index varies from 0.5 to 1.4 with domain length of 20$R$, Gnambode et al. [12] carried out a LES study of fully developed turbulent flows of power-law fluids in a cylindrical stationary pipe. The research of Gavrilova and Rudyak [13] [14] performed DNS of shear thinning fluid at two generalised Reynolds numbers 10000 and 20000, over the power law index 0.4-1. In 2016 [13] they focused on turbulent mean quantities by presented the distributions of components of Reynolds stress tensor, averaged viscosity, viscosity fluctuations, and measures of turbulent anisotropy. Gavrilova and Rudyak [14] reported a same study but this time they focused on the energy balance and the shear stresses by presented the distributions of the turbulent stress tensor components and the shear stress and turbulent kinetic energy balances. Their findings suggest that the increased flow index results in a pronounced enhancement in the effective viscosity, and, consequently, attenuation in the mechanism of the energy transport between the wall layer and the flow core and from axial to transverse component fluctuations, which also resulting in enhancement of the axial velocity fluctuations and attenuation of the fluctuations of the radial and tangential components. [15] examined the effect of the flow index parameter of power-law fluids in turbulent pipe flow are studied by means of direct numerical simulation at a friction Reynolds number 323, to understand the way in which shear thinning or thickening effects on the turbulent kinetic energy production, transport and dissipation in such flows.

The present study aims to critically examine the influence of the Non-Newtonian rheological and hydrodynamic behaviour on the turbulence main features, as well as to ascertain the accuracy and reliability of the laboratory code predicted results.

In addition to know to what extent the extended Smagorinsky model be able to characterise the scales motions especially in the wall vicinity. In this study, LES with an extended Smagorinsky model has been applied to fully developed turbulent flow of a shear thinning fluid with flow index of 0.75 in an axially pipe at a simulation's Reynolds number equals to 4000, the numeric resolution is $65^3$ gridpoints in streamwise, radial and spanwise directions respectively with a domain length of 20$R$. The current paper is organized as follows: governing equations and numerical procedure are presented in section 2. the effects of rotating wall of the pipe on the mean flow quantities and turbulent flow statistics (the root mean square (RMS) of the velocity fluctuations distribution, turbulent kinetic energy, turbulent Reynolds shear stress distributions and higher order statistics), and some contours of instantaneous velocity fluctuations are illustrated in section 3 and finally section 4 summarizes the results of this work and draws conclusions.

## 2. Governing equations and numerical procedure

### 2.1 Governing equations

The present study deals numerically with the fully developed of turbulent flow of a shear thinning fluid ($n$=0.75) fluids in an isothermal axially stationary pipe (Fig.1), by employing the LES technique at a simulation's Reynolds number of 4000. The filtered continuity and filtered momentum equations governing 3$D$ Non-Newtonian fluid are written in a cylindrical coordinate system and are made dimensionless using the centreline axial velocity of the analytical fully developed laminar profile, ($U_{CL}$=(3$n$+1).$U_b$/($n$+1)) as a reference velocity where $U_b$ is the average velocity, and the pipe radius $R$ as a reference length. The filtered equations can be written as follows:

$$\frac{\partial \overline{u}_i}{\partial x_i} = 0 \quad (1)$$

$$\frac{\partial \overline{u}_j}{\partial t} + \frac{\partial \overline{u}_i \overline{u}_j}{\partial x_i} = -\frac{d\overline{P}}{dx_j} + \frac{1}{Re_s}\frac{\partial}{\partial x_i}\left[\overline{\dot{\gamma}^{n-1}}\left(\frac{\partial \overline{u}_j}{\partial x_i} + \frac{\partial \overline{u}_i}{\partial x_j}\right)\right] + \frac{\partial \overline{\tau}_{ij}}{\partial x_i} \quad (2)$$

The overbar symbol ( ¯ ) denotes the filtering operation, and $Re_s$ is the simulation's Reynolds number which is defined as $Re_s = \rho U_{CL}^{2-n} R^n / K$.

The present modelled Non-Newtonian fluid is described by a power-law model with flow index (n<1) i.e. (a shear thinning fluid), the apparent viscosity is function of the shear rate magnitude $\dot{\gamma}$ as $\eta = K\dot{\gamma}^{n-1}$. subgrid stress tensor $\bar{\tau}_{ij}$ is associated to the strain rate tensor $\bar{S}_{ij}$ by $\bar{\tau}_{ij} = -2\nu_t \bar{S}_{ij}$, where the turbulent viscosity is computed by $\nu_t = C_s f_s (f_n \Delta)^2 \bar{S}_{ij}$ according to [11], where $\Delta$ is the computational filter, $C_s$ the model constant, $f_s$ the van Driest wall damping function, and $f_n$ the correction function for the change in viscosity.

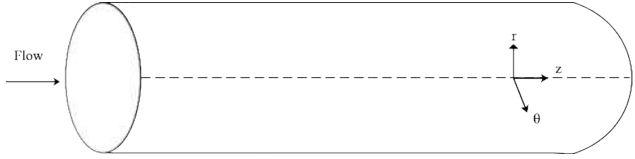

Fig.1 Schematic of the computational domain

## 2.2 Numerical procedure

The current laboratory code numerical integration is performed by a finite difference scheme, second-order accurate in space and time. Crank-Nicolson implicit scheme and a third-order Runge-Kutta explicit scheme were employed in order to treat the diffusive and convective terms, respectively, whereas a fractional step method was employed for the time integral. A periodic boundary condition has been assigned at the streamwise and spanwise directions, No-slip boundary condition has also imposed to the pipe wall. The table.1 summarises the grid resolution details and the mean flow quantities.

Tab.1: parameters of present simulations

| Parameter | n=0.75 | Parameter | n=0.75 |
|---|---|---|---|
| $\Delta z^+$ | 69.89 | $\eta_{d,w}$ | 0.6334 |
| $r\Delta\theta^+$ | 21.95 | $Re_{cr}$ | 2250 |
| $\Delta r^+_{min}$ | 0.0506 | $Re_s$ | 4000 |
| $\Delta r^+_{max}$ | 12.4 | $Re_{MR}$ | 4873.2 |
| $U_b/U_{cl}$ | 0.5329 | $Re_g$ | 7135.8 |
| $U_c/U_{cl}$ | 0.6863 | $Re_\tau$ | 227.16 |
| $U_\tau/U_{cl}\,10^1$ | 0.3393 | $Y_1^+$ | 0.0239 |
| $\dot{\gamma}_{d,w}$ | 6.6818 | $f\cdot 10^3$ | 9.2099 |

## 3. Results and discussion

The statistical turbulence quantities of the turbulent flow such as the turbulent kinetic energy, root mean square (RMS) of the velocity fluctuations, turbulent Reynolds shear stress, higher order statistics and contours of velocity fluctuations have been critically analysed and discussed here, in order to provide a deeper insight into the influence of the Non-Newtonian rheological and hydrodynamic behaviour on the turbulence main features, as well as to ascertain the accuracy and reliability of the laboratory code predicted results. For the validation purpose, the present LES predictions compared reasonably well with a Newtonian DNS data performed by [17] at $Re$=5500 and a shear thinning fluid ($n$=0.75) DNS data performed by [9] at $Re_{MR}$=3935. The Fig.2a compares the RMS of the mean axial, radial and tangential velocity of the Newtonian fluid with [17], while that the Fig.2b compare the radial axial turbulent Reynolds shear stress of the shear thinning fluid ($n$=0.75) [9], along the pipe radius at a simulation's Reynolds number of 4000. As shown in Fig.2a the RMS of the streamwise velocity fluctuation is in an excellent agreement with the DNS data of [17] over the pipe radius, whereas the two others components of the RMS are slightly underestimated in comparison with those of [17], this little discrepancy between them may be due to the difference in the numerical methods and Reynolds number.

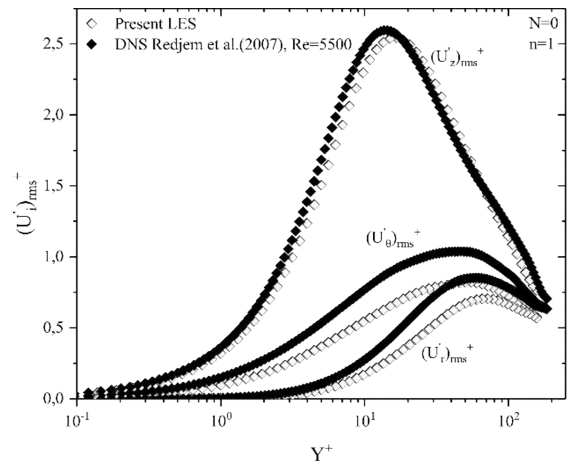

Fig.2a: RMS of velocity fluctuations

As seen in Fig.2(b), it is apparent that no significant noteworthy differences were found between the predicted profile and that obtained by [9], overall, these predicted profiles are in qualitative agreement with those obtained numerically by DNS.

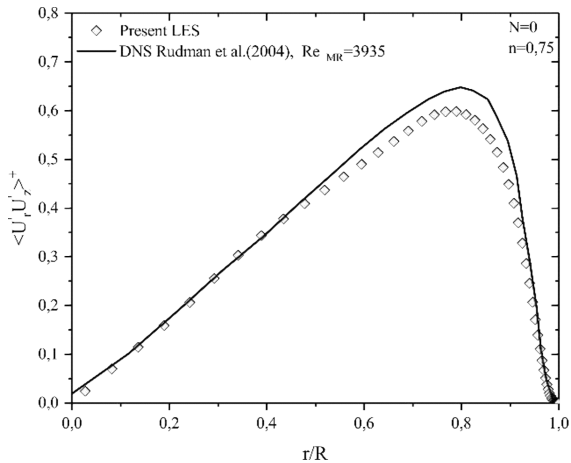

Fig.2b: Turbulent Reynolds stress

### 3.1 Root mean squares of velocity fluctuations

The following paragraphs offer a brief overview of the velocity fluctuations distributions over the pipe in addition to the generation and transport mechanism of the turbulent energy in the flow regions. The Fig.3 presents the root mean square (RMS) distribution of the axial, radial and tangential velocity fluctuations of the shear thinning ($n$=0.75) and Newtonian fluids ($n$=1), scaled by the friction velocity $U_\tau = \sqrt{\tau_w/\rho}$ along the pipe radius, versus the distance from the wall in wall units $Y^+$ at a simulation's Reynolds number of 4000. As seen in Fig.3, the RMS of the axial, radial and tangential velocity fluctuations of the shear thinning and Newtonian fluids are nearly characterised by the same trend along the radial direction with a little discrepancy between them out of the viscous sublayer ($Y^+$=5), the axial profile of the shear thinning lies higher than the Newtonian fluid. On the contrary, the radial and tangential profiles of the shear thinning lie under the Newtonian one. An overall increased in the axial fluctuations of the shear thinning fluid occurs over the pipe radius, which results in an attenuation in the radial and tangential fluctuations velocity. As observed in Fig.3, the axial, radial and tangential velocity fluctuations profiles of the shear thinning fluid ($n$=0.75) coincide with the profiles of the Newtonian fluid in the wall vicinity, these profiles deviate noticeably from each other only beyond $Y^+$= 7 towards the core region with the wall distance, particularly in the buffer layer (5 ≤ $Y^+$ ≤ 30) and logarithmic layer (30 ≤ $Y^+$ ≤ 200), this discrepancy is more pronounced in the streamwise profile compared to the radial and spanwise ones as shown in the zoomed view. In addition, the peak value of the shear thinning fluid shifts slightly towards the core region and raises noticeably compared to the other components, which indicates that the intensity of the axial fluctuations of the shear thinning fluid is more pronounced than that of the Newtonian one.

The RMS of the axial, radial and tangential velocity fluctuations of the shear thinning fluid ($n$=0.75) are nearly zero in the vicinity of the wall, these profiles enhance gradually away from the wall and attain their maximum values, which means that the axial velocity fluctuations are generated in the wall vicinity and are propagated far away from the wall towards the core region: the fluctuations of the axial velocity component generate in the wall vicinity and transfer to the radial and tangential components, these results are consistent with that of [13]. At larger distances from the wall $Y^+$>20, these velocity fluctuations profiles begin to decrease gradually towards the core region with the distances from the wall, which indicates that the velocity fluctuations start to vanish over the remaining regions.

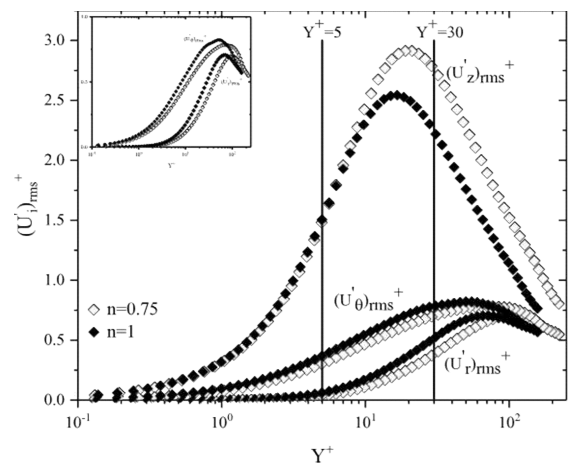

Fig.3 RMS of velocity fluctuations

## 3.2 Turbulent Kinetic Energy

The Figure.4 presents the distributions of the kinetic energy of turbulent fluctuations for the shear thinning ($n$=0.75) and Newtonian fluids along the pipe radius versus the distance from the wall in wall units $Y^+$, at simulation's Reynolds number of 4000. The shear thinning fluid and the Newtonian fluid have the same trend over the pipe radius with a minor discrepancy, where the Newtonian profile lies above that of the shear thinning fluid. For the shear thinning fluid ($n$=0.75), the turbulent kinetic energy is nearly constant and equals to zero in the vicinity of the wall ($Y^+$=1), as a result of the fluctuations absence in this region. Beyond $Y^+$=1, the turbulent kinetic energy enhances gradually away from the wall towards the core region with the distance from the wall up to $Y^+$=20 where exhibits its maximum value (0.018), which is corresponding to the peak of the axial velocity fluctuations profile. Indeed, this enhancement due to the axial velocity fluctuations generation near the wall, in addition to the energy transfer from the axial fluctuations to the two other components. At the larger distances from the wall ($Y^+$=20), the kinetic energy of the turbulent fluctuations drops monotonically in the core region as a consequence of the decreased in the velocity fluctuations in this region.

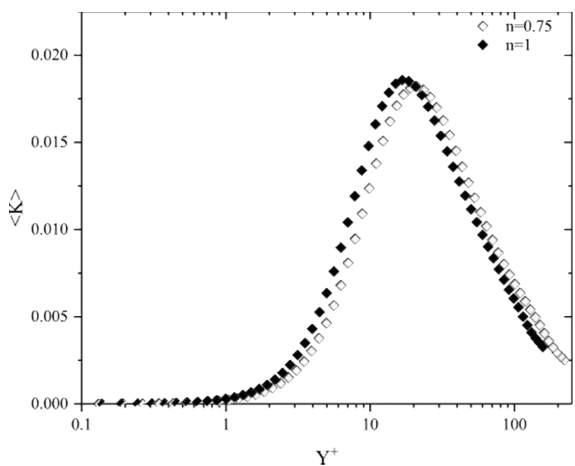

Fig.4 Turbulent kinetic energy

## 3.3 Turbulent Reynolds shear stress

Fig.5 displays the turbulent Reynolds shear stress profile of a shear thinning ($n$=0.75) and Newtonian fluids, scaled by friction velocity along the pipe radius versus the distance from the wall in wall units at $Re_s$=4000. As seen in Fig.5, In the vicinity of the wall (viscous sublayer $Y^+$=5), the profile of the axial radial turbulent shear stress is nearly constant and equals to a zero, beyond $Y^+$=3 the turbulent shear stress profile increases gradually away from the wall towards the flow core with the wall distance up to $Y^+$=70, this enhancement is due to the generation of the axial velocity fluctuations in the wall vicinity in addition to transport of these fluctuations to the radial and tangential components from the wall towards the core region. The turbulent Reynolds shear stress profile decreases after reaching the peak value (0.6) at a wall distance of 70 wall unites this profile falls off rapidly to zero value in the core region, which indicates the fluctuations of the velocity are vanished in this region.

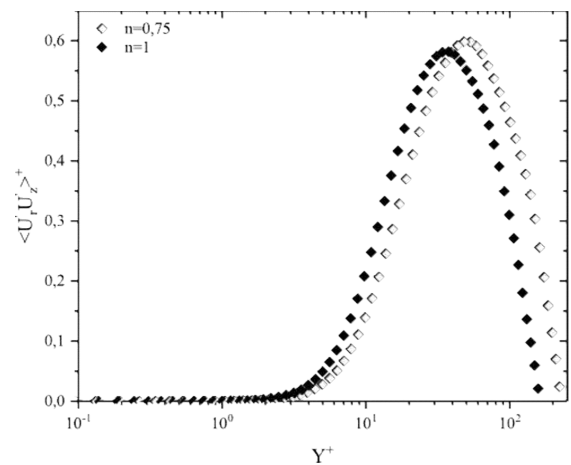

Fig.5 Turbulent Reynolds stress

## 3.4 Higher order statistics

In order to provide a further indication of the intermittent character of the wall region, the turbulent higher order statistics of the shear thinning ($n$=0.75) and the Newtonian fluids are presented and discussed in the following paragraphs. Fig.6 and Fig.7 illustrate respectively the predicted skewness and flatness distributions of the streamwise velocity fluctuation along the pipe radius, versus the distance from the wall in wall units $Y^+$ a simulation's Reynolds number of 4000. Apparently, there is a clear trend of monotonically decreasing in the skewness and flatness coefficients along the radial direction from the wall to the pipe centre with the distance from the wall $Y^+$, it is evident that the profiles of the shear thinning fluid almost coincide with that of the Newtonian one along the pipe radius. For the shear thinning fluid, the skewness profile seems the higher values in the wall vicinity then tends rapidly to the Gaussian value ($S(U'_z) = 0$) as shown in Fig.6.

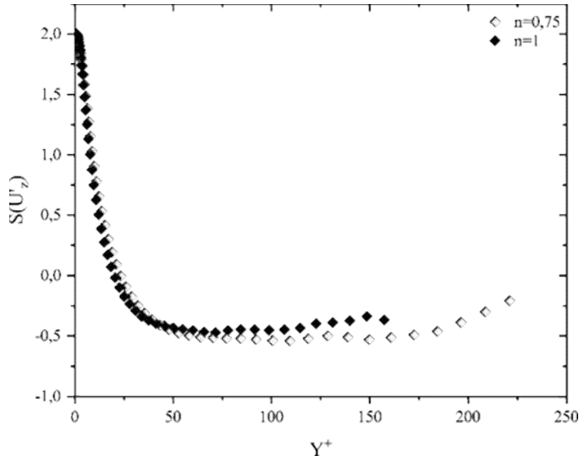

Fig.6 Skewness of temperature fluctuations

The large positive skewness in the viscous sublayer indicates that large positive values of the mean axial velocity components are predominated rather than the large negative values in this region. Moreover, the flatness profile exhibits the higher values near the pipe wall where these values are ascribed to the strong sweep events near by the pipe wall, which reflects that the intermittent behaviour near the wall is more pronounced than in the pipe centre. The flatness profile drops rapidly away from the wall towards the pipe centre with the wall distance, this profile goes to the Gaussian value ($F(U'_z) = 3$) as seen in Fig.7.

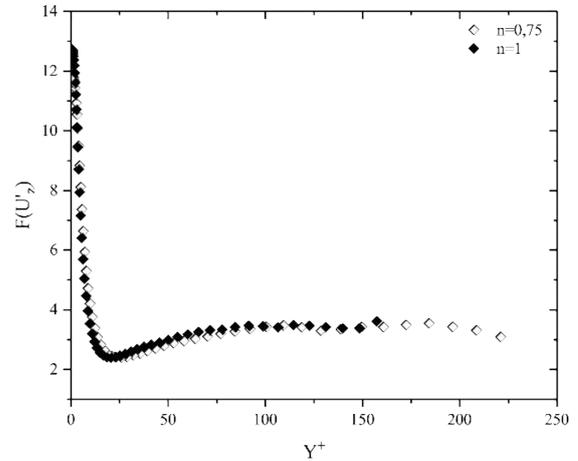

Fig.7 Flatness of axial velocity fluctuations

### 3.5 *Visualisations*

The contour lines of a snapshot of the streamwise and radial velocity fluctuations over a cross section in *r*-θ plane of the shear thinning fluid (*n*=0.75) are visualised in Fig.8, where the solid lines refer to the positive value whereas the dash ones refer to the negative values. As appeared near wall region the positives value of the velocity fluctuations is dominated by high intensity fluctuations Fig.8(a) which indicates the axial velocity fluctuation are generated in this region.

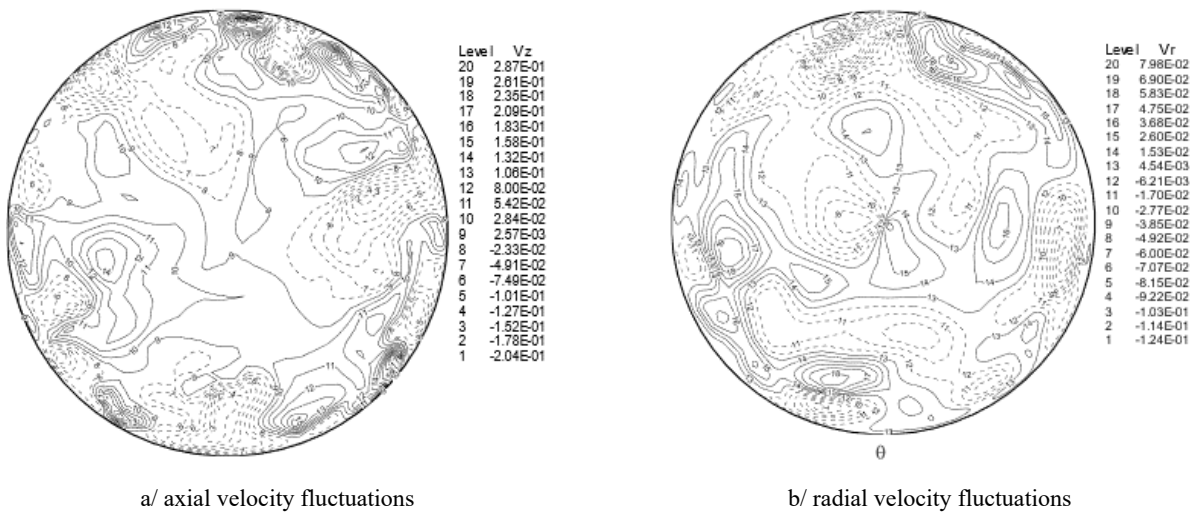

a/ axial velocity fluctuations    b/ radial velocity fluctuations

Fig.8 Contour lines of the velocity fluctuations

The turbulent structures intensity decreases progressively from the wall towards the core region, meaning that these fluctuations are started to vanish far away from the pipe wall (Fig.3). It is worth noting that the turbulent energy of streamwise fluctuations is transferred to two other components, which resulted in a pronounced enhancement in the radial and spanwise velocity fluctuations over the core region, see Fig.8(b).

**Conclusion**

The present paper reports on numerical computations of a fully developed turbulent flow of a shear thinning fluid with flow index of 0.75 in an isothermal axially stationary pipe flow, by applying a large eddy simulation with an extended Smagorinsky model. The purpose of the current study was to evaluate the influence of the Non-Newtonian rheological and hydrodynamic behaviour on the turbulence main features, as well as to ascertain the accuracy and reliability of the laboratory code predicted results. This accuracy has been ascertained by comparing the predicted results with those available in the literature. Many statistical quantities were obtained in the present study such as (root mean square (RMS) of the velocity fluctuations distributions, turbulent kinetic energy, turbulent Reynolds shear stress and higher order statistics) and some contours of instantaneous velocity fluctuations.

The mean axial velocity fluctuations are generated in the wall vicinity and transfer to the radial and spanwise components, where the axial velocity fluctuations of the shear thinning fluid are high than the Newtonian one especially in the logarithmic layer, which leads to enhancement in the kinetic energy of turbulent fluctuations along the radial direction. On the contrary, the radial and tangential velocity fluctuations of the shear thinning are less than the Newtonian one in this region, this means that the energy transport from the axial fluctuations to the other fluctuations components between the flow layers are suppressed. Moreover, the skewness and flatness seem nearly independent of the flow index.